\documentclass[aps, prb, 10pt, twocolumn, superscriptaddress]{revtex4-2}
\usepackage[plainpages=false, colorlinks=true, linkcolor=blue, citecolor=blue, urlcolor=blue]{hyperref}
\usepackage[english]{babel}
\usepackage{graphicx, comment}
\usepackage[charter]{mathdesign}
\usepackage{amsmath}
\usepackage{slashed}
\usepackage{bm}
\usepackage{listings}
\usepackage{hyperref}
\usepackage{float}
\begin{document}
\title{Topological analog of the magnetic bit within the Su-Schrieffer-Heeger-Holstein model}
\author{Xinyuan Xu}
\affiliation{D\'epartement de physique and Institut quantique, Universit\'e de Sherbrooke, Sherbrooke, Qu\'ebec, Canada J1K 2R1}
\author{David S\'en\'echal}
\affiliation{D\'epartement de physique and Institut quantique, Universit\'e de Sherbrooke, Sherbrooke, Qu\'ebec, Canada J1K 2R1}
\author{Ion Garate}
\affiliation{D\'epartement de physique and Institut quantique, Universit\'e de Sherbrooke, Sherbrooke, Qu\'ebec, Canada J1K 2R1}

\date{\today}            

\begin{abstract}

In magnetic memories, the state of a ferromagnet is encoded in the orientation of its magnetization. The energy of the system is minimized when the magnetization is parallel or antiparallel to a preferred (easy) axis. These two stable directions define the logical bit. Under an external perturbation, the direction of magnetization can be controllably reversed and thus the bit flipped.
Here, we theoretically design a topological analogue of the magnetic bit in the Su-Schrieffer-Heeger (SSH)-Holstein model, where we show that a transient external perturbation can lead to a permanent change in the electronic band topology.

\end{abstract}
\maketitle

\newcommand{\tr}{\mathop{\mathrm{tr}}}
\newcommand{\Tr}{\mathop{\mathrm{Tr}}}

\section{Introduction}

In the past 15 years, there has been a flurry of research activity in the area of topological condensed matter physics (see e.g. Refs.~[\onlinecite{franz2013topological, bernevig2013topological, vanderbilt2018berry, armitage2018weyl, moessner2021topological, bernevig2022progress, wieder2022topological, das2023search}] and references therein). 
One of the research frontiers in the field is to develop the ability to control and alter the values of topological invariants through the application of external perturbations~\cite{oka2019floquet,de2021colloquium}. Such control could be a precursor of topological transistors and memories.  

There exist conceptually different approaches to switching the value of topological invariants. One of them is to induce a reversible topological phase transition in thermodynamic equilibrium through the application of pressure~\cite{xi2013signatures} or of a static electric field~\cite{bampoulis2023quantum}. 
A second approach relies on Floquet engineering: a topologically trivial material is driven out of equilibrium by a time-periodic perturbation, such that the dressed electronic bands become topologically nontrivial~\cite{rudner2020band}.
One limitation of the preceding two approaches is that the induced change in the electronic topology is transient: once the external perturbation is removed, the system goes back to its initial state.

A third approach, pursued more recently, consists in utilizing light pulses to induce a transition between two stable (or metastable) states of different electronic band topology~\cite{sie2019ultrafast, zhang2019light, zhou2021terahertz}. This approach has the advantage of allowing a permanent change in electronic topology under the transient action of an external perturbation. One challenge of this approach is that, thus far, it has been implemented only in materials whose crystal structure must be modified in order to switch the electronic topological invariant. The intense light pulses required to attain such change in structure are hardly compatible with scalable transistor or memory architectures. In addition, the two topologically distinct phases in those materials are both electrically conducting (a Weyl semimetal and a topologically trivial semimetal), which makes it more difficult to distinguish them and use them for applications.

In the present work, we theoretically conceive a system in which a small transient external perturbation can lead to a permanent switch between two topologically distinct phases of matter that are electrically insulating. 
Our proposal is inspired by an analogy with the magnetic bit.
In magnetic memories~\cite{tang2010magnetic}, the state of a magnetic material is encoded in the orientation of its magnetization. The energy of the system is minimized when the magnetization is parallel or antiparallel to a preferred axis. These two stable directions define the magnetic bit ($0$ and $1$). Under an external perturbation, the direction of magnetization can be reversed following the Landau-Lifshitz-Gilbert equation~\cite{spintronics} and thus the bit flipped. To do so, the external perturbation must overcome the magnetic anisotropy energy, which is small compared to the energy scale associated to the magnitude of the magnetization. Accordingly, relatively weak perturbations suffice to flip the magnetic bit.
At the same time, the magnetic anisotropy energy ensures that the magnetization will retain its new stable state, with robustness under thermal and quantum fluctuations, when the perturbation is turned off.  

In our work, the roles of $0$ and $1$ are played by two energetically degenerate ground states having different values of a topological invariant. 
This idea is similar in spirit to the third approach described above, while differing from it in a number of ways. First, we do not require strong external perturbations. Second, the phase transition is between two insulating phases of matter. Third, the topological invariant is encoded not in a discrete crystal structure, but instead in an order parameter that can be continuously changed. 
Admittedly, a straightforward example of such encoding is realized in an anomalous Hall insulator, wherein the reversal of the magnetic order parameter results in a change of the electronic topological invariant (Chern number).
Yet, this change occurs between two topologically nontrivial states (see also  Ref.~\cite{claassen2019universal} for a related switch in the superconducting context). 
Here, we are instead interested in attaining a switch between a topologically trivial phase and a topologically nontrivial phase, both of them being insulating in the bulk.

We find that the simplest system realizing the preceding attributes is the SSH-Holstein model at half-filling, with its parameters tuned close to a phase transition between a bond density-wave and a site density-wave. 
This is a situation that does not necessarily describe a real material; it remains an open question whether the ideas exposed in it will be eventually realized in experiments. 
The model consists of a one-dimensional chain of orbitals with SSH-type (inter-site) and Holstein-type (on-site) electron-phonon coupling. When the SSH coupling dominates, the ground state displays an instability towards a bond-density-wave. This bond-density-wave has two topologically distinct but energetically identical ground states (Fig.~\ref{fig:cartoon}b), which are the analogues of the two states of a magnetic bit. 
Shifting the order parameter to a site density-wave  (Fig.~\ref{fig:cartoon}a) has an energy cost that plays a role analogous to the magnetic anisotropy energy in magnetic memories.

The rest of this work is organized as follows. In Sec.~\ref{eq:sshh}, we review the mean-field phase diagram and associated topological invariants of the SSH-Holstein model.
In Sec.~\ref{sec:op}, we describe the mean-field order parameter dynamics under external perturbations. 
We use the path-integral formalism to derive an equation of motion of the density-wave order parameter, and we identify parameter regimes in which the topological analogue of magnetization reversal is realized.
In Sec.~\ref{sec:jj}, we  highlight connections with the dynamics of the superconducting order parameter in an unconventional Josephson junction. 
In Sec.~\ref{sec:switch}, we provide a proof of principle for the topological analogue of the magnetic bit.
The main text of the paper concludes in Sec.~\ref{sec:conc} with a summary of our results and a  comparison with the existing literature on the dynamics of incommensurate charge density waves.
Some technical points are relegated to the Appendix.
We take $\hbar\equiv 1$ until Sec.~\ref{sec:mot}, and restore the $\hbar$ factors afterwards.

\section{The SSH-Holstein model}
\label{eq:sshh}

\begin{figure}
\includegraphics[width=\hsize]{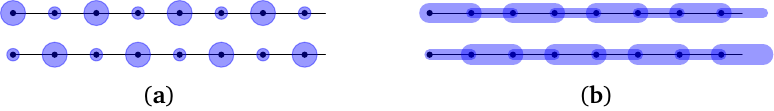}
\caption{Schematic representation of the mean-field ground states for the Holstein model (panel (a)) and the SSH model (panel (b)), with the electron density depicted in blue.
In each panel, the two degenerate ground states (differing by a sign of the order parameter) are displayed.
 \label{fig:cartoon}}
\end{figure}
The SSH-Holstein model describes spinless fermions moving in one dimension, with a hopping amplitude and an on-site energy modulated by two distinct types of phonons. The corresponding Hamiltonian reads
\begin{align}
\label{eq:SSH-H}
H&=t_{0}\sum_{j}(c_{j}^{\dagger}c_{j+1}+\mathrm{H.c.})-\alpha_{S}\sum_{j}(q_{j}-q_{j+1})(c_{j}^{\dagger}c_{j+1}+\mathrm{H.c.})\notag\\
&+\frac{1}{2}\sum_{j} \left[m_S \dot{q}_j^2+K_{S}(q_{j}-q_{j+1})^{2}\right]\notag\\
&-\alpha_{H}\sum_{j}Q_{j} c_{j}^{\dagger}c_{j}+\frac{1}{2}\sum_{j}\left(m_{H}\dot{Q_{j}}^{2}+K_{H}Q_{j}^{2}\right)~,
\end{align}
where $j$ is the site index, $t_{0}$ is the hopping amplitude between nearest neighbors (assumed to be real), $q_{j}-q_{j+1}$ is the bond displacement (SSH phonon), and $Q_{j}$ is the intra-site displacement (Holstein phonon). 
Also, $\alpha_i$, $K_i$ and $m_i$ are, respectively, the electron-phonon couplings, the spring constants and atomic masses associated to SSH ($i=S$) and Holstein ($i=H$) phonons. 

When $\alpha_H=0$, Eq.~(\ref{eq:SSH-H}) becomes the SSH model~\cite{su1979solitons}.
At half-filling (one electron per two sites), this system undergoes a Peierls instability towards a bond density-wave (Fig.~\ref{fig:cartoon}b).
In the mean-field approximation~\cite{fradkin1983phase}, the density-wave is described by the staggered displacement field 
\begin{equation}
\label{eq:qj}
 q_{j}=(-1)^{j}\Delta_S, 
\end{equation}
where $\Delta_S$ is the SSH order parameter. 
Substituting Eq.~(\ref{eq:qj}) in Eq.~(\ref{eq:SSH-H}), assuming a static and uniform order parameter $\Delta_{0S}$, adopting the continuum approximation (see Sec.~\ref{sec:cont} below) and minimizing the energy with respect to $\Delta_{0S}$, we find
\begin{equation}
\label{eq:Delta0x}
\Delta_{0S} \simeq \pm \frac{a t_0 \Lambda}{\alpha_S} e^{-1/\lambda_S},
\end{equation}
where 
$a$ is the lattice constant (distance between nearest neighbors), $\Lambda\sim 1/a$ is a UV wave vector cutoff, and 
\begin{equation}
\label{eq:KStilde}
\lambda_S= \frac{2 \alpha_S^2}{\pi t_0 K_S}
\end{equation}
is a dimensionless coupling constant between electrons and SSH phonons ($\lambda_S\ll 1$ in the continuum approximation).

Although the ground state energy is independent  of the sign of $\Delta_{0S}$,  the electronic band topology is sensitive to it. 
For one sign, the electronic Berry phase is $\pi\,({\rm mod}\,2\pi)$; for the other sign, the electronic Berry phase is $0\,({\rm mod}\,2\pi)$. 
In the first case, but not in the second, a finite-size chain hosts zero-energy electronic modes at chain ends~\cite{franz2013topological}. 
Domain walls between regions of opposite sign of $\Delta_{0S}$ likewise host zero-energy electronic modes. These modes are protected by chiral symmetry.

When $\alpha_S=0$, Eq.~(\ref{eq:SSH-H}) becomes the Holstein  model~\cite{holstein1959studies}.
At half filling, this system undergoes a Peierls instability towards a site-density-wave  (Fig.~\ref{fig:cartoon}a),
\begin{equation}
\label{eq:Qj}
Q_{j}= (-1)^{j} \Delta_H, 
\end{equation}
where $\Delta_H$ is the Holstein order parameter.  
Assuming a static and uniform order parameter $\Delta_{0H}$, the minimization of the mean-field energy in the continuum approximation gives
\begin{equation}
\label{eq:Delta0z}
\Delta_{0H} \simeq \pm \frac{4 a t_0 \Lambda}{\alpha_H} e^{-1/\lambda_H},
\end{equation}
where 
\begin{equation}
\label{eq:KHtilde}
\lambda_H= \frac{\alpha_H^2}{2\pi t_0 K_H}
\end{equation}
is a dimensionless coupling constant between electrons and Holstein phonons  ($\lambda_H\ll 1$ in the continuum approximation).

\begin{figure*}[t]
\includegraphics[width=1\hsize]{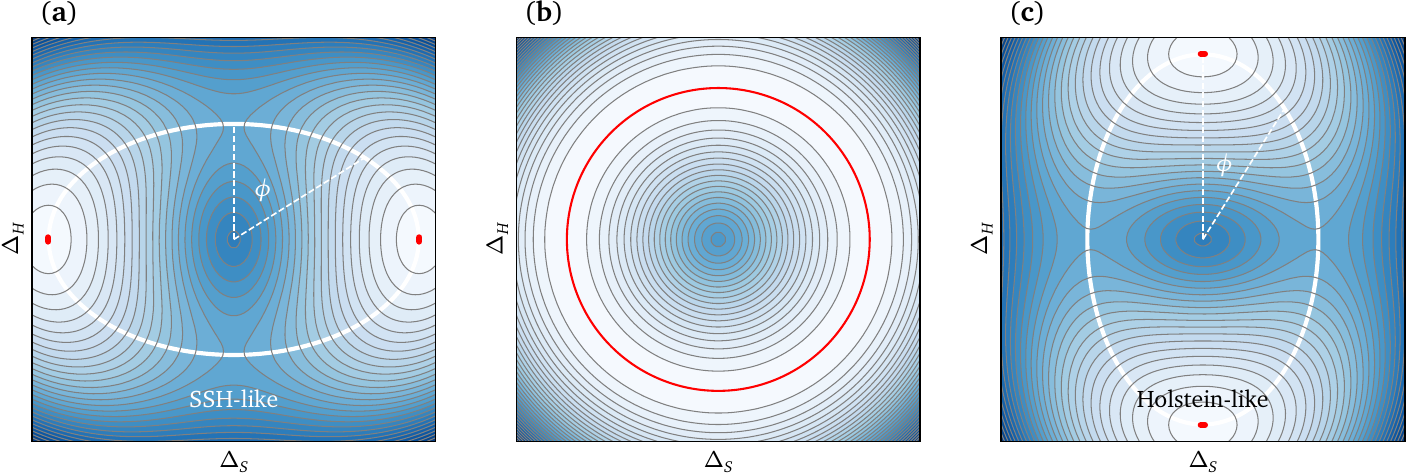}
\caption{Mean-field energy for the SSH-Holstein model calculated in the continuum approximation, as a function of the order parameter $\boldsymbol{\Delta}=(\Delta_S, \Delta_H)$.
Darker blue denotes higher energy. Energy minima are indicated in red.
(a) When $\lambda_S>\lambda_H$  (defined in Eqs.~(\ref{eq:KStilde}) and (\ref{eq:KHtilde})), the ground state is a doubly degenerate bond density-wave.
The white line indicates the lowest-energy path for switching the order parameter between the two degenerate ground states. This is analogous to the path followed in the magnetization reversal of a two dimensional ferromagnet with easy axis anisotropy. In the main text, we study the dynamics of the order parameter angle $\phi$ as a function of external perturbations. 
(b) When $\lambda_S=\lambda_H$, the ground state is infinitely degenerate (red line). In the lattice model, the phase boundary between the SSH-like and Holstein-like charge density-waves is slightly away from $\lambda_S=\lambda_H$.
(c) When  $\lambda_S<\lambda_H$ , the ground state is a doubly degenerate site density-wave. 
\label{fig:energy}}
\end{figure*}

Like in the SSH model, the ground state of the Holstein model is doubly degenerate because the system's energy is independent of the sign of $\Delta_{0H}$.
Contrary to the SSH model, the two ground states of the Holstein model are not distinct from the point of view of the Berry phase, which varies continuously with the parameters of the Hamiltonian.
Yet, the two signs of $\Delta_{0H}$ are topologically distinct due to a nonsymmorphic chiral symmetry~\cite{shiozaki2015}.
Unlike the chiral symmetry of the SSH model, the nonsymmorphic chiral symmetry of the Holstein model is broken at the ends of a finite-size chain, as well as at a sharp domain wall between regions of opposite sign of $\Delta_{0H}$.  
As a result, zero-energy modes are not guaranteed at interfaces of different topological invariant~\cite{allen2022}. 
 
When both $\alpha_S$ and $\alpha_H$ are nonzero, the nature of the mean-field ground state in the continuum approximation depends on the ratio $\lambda_S/\lambda_H$:
(1) when $\lambda_S>\lambda_H$, the ground state harbors a SSH-like (bond density-wave) order parameter (Eq.~(\ref{eq:Delta0x})); (2) when $\lambda_S<\lambda_H$, the ground state displays a Holstein-like (site density-wave) order parameter (Eq.~(\ref{eq:Delta0z})); (3) when $\lambda_S=\lambda_H$, the mean-field order parameter is infinitely degenerate with
\begin{equation}
\left(\Delta_{0S},\Delta_{0H}\right) = 4 a t_0 \Lambda e^{-1/\lambda_H}\left(\frac{1}{4\alpha_S} \sin\varphi, \frac{1}{\alpha_H} \cos\varphi\right),
\end{equation}
for any $\varphi\in[0,2\pi)$. 
In this case, the charge density-wave is said to be incommensurate.
In comparison, the charge density-wave for $\lambda_S\neq \lambda_H$ is named as commensurate. 
Figure~\ref{fig:energy} illustrates the mean-field energy landscape for the SSH-Holstein model.

The preceding observations are based on the mean-field approximation, which is expected to be accurate in the adiabatic regime~\cite{bakrim2007quantum}, where the phonon frequencies 
are small compared to the mean-field energy gaps $(\alpha_S |\Delta_{0S}|, \alpha_H |\Delta_{0H}|)$. In the nonadiabatic regime, the density-waves are less robust and metallic phases may also emerge (see e.g. Refs.~[\onlinecite{hohenadler2016interplay, hohenadler2018density}] for a recent numerical study on the spinful SSH-Holstein model). Hereafter, we will limit ourselves to the adiabatic case.

In order to realize a topological analogue of the magnetic bit, we place ourselves in the regime $\lambda_S>\lambda_H$. 
In this regime, the two ground states have topologically robust Berry phases, one of them being zero (trivial) and the other $\pi$ (topological).
As illustrated in Fig.~\ref{fig:energy}a, the situation is reminiscent to that of a two-dimensional ferromagnet with easy-axis anisotropy: the two preferred (anti-parallel) orientations of the magnetization can be mapped to the two signs of the order parameter $\Delta_{0S}$.

In a ferromagnet with easy axis anisotropy, the dynamics of the magnetization and its possible reversal in the presence of external perturbations are described by the Landau-Lifshitz-Gilbert (LLG) equation~\cite{spintronics}.
In the next section, we will search for a counterpart of the LLG equation in order to describe the reversal of electronic band topology in the SSH-Holstein model.
Such reversal will take place following the low-energy path indicated in white in Fig. \ref{fig:energy}a. 
The probability for direct quantum tunneling of the order parameter through the energy barrier at $(\Delta_S, \Delta_H)= (0,0)$ will be neglected.

\section{Order parameter dynamics}
\label{sec:op}

Thus far we have discussed the mean-field ground state of the SSH-Holstein model and we have described the presence of topologically distinct ground states. 
 In this section, we determine the dynamics of fluctuations around the mean-field ground state.

\subsection{Continuum model approximation}
\label{sec:cont}

We begin by applying the continuum approximation to the SSH-Holstein model, which will facilitate the calculations that follow. The continuum model is motivated by the fact that, for half-filling, low energy excitations are concentrated in the vicinity of the Fermi wave vectors $\pm k_F$ with $k_F=\pi/2a$~\cite{giamarchi2003quantum}. This allows to write 
\begin{equation}
    \frac{c_{j}}{\sqrt{a}}\simeq \Psi_{L}(x)e^{ik_{F}x}+\Psi_{R}(x)e^{-ik_{F}x},
    \label{31}
\end{equation}
where $\Psi_{L}$ ($\Psi_R$) is the field operator for the left-moving (right-moving) fermions, varying slowly as a function of $x$. Substituting Eqs.~(\ref{eq:qj}), (\ref{eq:Qj}) and (\ref{31}) into Eq.~(\ref{eq:SSH-H}) and keeping only terms that vary slowly at the scale of $a$, the mean-field Hamiltonian of the SSH-Holstein model becomes 
\begin{align}
H &=\int dx \Psi^{\dagger}h\Psi+\frac{m_{S}}{2a} \int dx\left(\dot{\Delta}_S^{2}+\omega_{S}^{2}\Delta_S^{2}\right)\notag\\
&+\frac{m_{H}}{2a}\int dx\left(\dot{\Delta}_H^{2}+\omega_{H}^{2}\Delta_H^{2}\right),
\label{eq:H1}
\end{align}
 where $\omega_H=(K_H/m_H)^{1/2}$ and $\omega_S=(4 K_S/m_S)^{1/2}$ are frequencies of Holstein and SSH phonons,
$\Psi=(\Psi_{L},\Psi_{R})$ is a two-component spinor and
\begin{align}h&= -iv\sigma^{z}\partial_{x}-4\alpha_S\Delta_S\sigma^{y}+\alpha_{H}\Delta_H\sigma^{x}
\label{32}
\end{align}    
is the electronic Hamiltonian for a 1D Dirac fermion \cite{goldstone1981fractional} with scalar and pseudoscalar masses $(-4\alpha_S\Delta_S, \alpha_H\Delta_H)$ and velocity $v=2at_{0}$, $\sigma_i$ being the Pauli matrices in the $(L,R)$ space. 
When $\Delta_S$ and $\Delta_H$ are treated as known parameters (rather than order parameters with internal dynamics), the electronic part of the SSH-Holstein model is also widely known as the Rice-Mele model \cite{cayssol2021topological}.
Below, we will study the dynamics of $\Delta_S$ and $\Delta_H$ in response to external perturbations.

\subsection{Perturbations}

We will focus on external perturbations that act on electrons. In the presence of such perturbations, the Hamiltonian in Eq.~(\ref{eq:H1}) acquires an additional term
\begin{equation}
\delta H=\int dx \Psi^{\dagger}\delta h\Psi=\int dx \Psi^{\dagger}\left(\mathbf{B}\cdot\boldsymbol{\sigma}+B_{0}\right)\Psi,
\label{33}
\end{equation}
where ${\bf B}=(B_x, B_y, B_z)$.
Physically, $B_0$ and $B_z$ can be interpreted as scalar and vector electromagnetic potentials, respectively, because $B_{0}$ couples to the charge density $\Psi^\dagger\Psi$ and $B_{z}$ enters the Hamiltonian density as $\Psi^\dagger(-iv\partial_{x}+B_{z})\sigma^{z}\Psi$.
Likewise, $B_x$ and $B_y$ can be regarded as external influences on the Dirac masses in Eq.~(\ref{32}).

In general, perturbations that lead to electron backscattering result in nonzero $B_x$ and $B_y$. 
Static disorder is an example of such a perturbation. There are, in addition, other sources of $B_x$ and $B_y$. 
For example, let us consider a spatial modulation of the onsite potential,
\begin{equation}
\label{eq:Vj}
    \sum_{j}V_{j}c_{j}^{\dagger}c_{j}.
\end{equation}
Using Eq.~(\ref{31}) and transferring $V(x)$ to Fourier space as $V(x)=\sum_{k}v_{k}e^{ik x}$, Eq.~(\ref{eq:Vj}) becomes proportional to
\begin{equation}
   \int dx \Psi^{\dagger}[v_{0}+{\rm Re}({v_{2k_{F}}})\sigma^{x}-{\rm Im}({v_{2k_{F}}})\sigma^{y}]\Psi.
 \end{equation}
Thus, a modulation of the onsite potential produces the perturbations $B_0$, $B_x$ and $B_y$. If the modulation is symmetric in space ($V(x)=V(-x)$) and has a zero mean ($\int V(x) dx =0$), then only $B_x$ is present.
Similarly, in the case of modulated strain, 
\begin{equation}
  \sum_{j}V_{j}(c^{\dagger}_{j}c_{j+1}+{\rm H.c.})\propto  {\rm Re}({v_{2k_{F}}})\int dx \Psi^{\dagger} \sigma^{y}\Psi,
\end{equation}
we get a nonzero $B_y$.
A recent theoretical study \cite{mori2023shapiro} indicates that $B_y$ could also be produced when the system containing impurities is driven by ultrasound.

\subsection{Effective action for the order parameters}

A convenient and standard way to obtain the order parameter dynamics is through the minimization of an effective action. To obtain such an effective action, we begin by writing the partition function of the system in the imaginary-time path integral formalism~\cite{altland2010condensed}:
\begin{equation}
    Z=\int D(\Psi^{\dagger},\Psi,\Delta_S,\Delta_H) \ e^{-S[\Psi^{\dagger},\Psi,\Delta_S,\Delta_H ]},
\end{equation}
 where
 \begin{align}
S&=\int d\tau dx\left[\Psi^{\dagger}\partial_{\tau}\Psi-\mu\Psi^{\dagger}\Psi\right]+\int d\tau \left(H+\delta H\right)\label{39}
\end{align}
 is the Euclidean action of the system, $\mu$ is the chemical potential (we take $\mu=0$ below, corresponding to half-filling), and $\tau=it$ is the imaginary time. The full action in Eq.~(\ref{39}) contains both fermion and phonon (order parameter) fields. However, we are interested only in the dynamics of the order parameters. 
 Thus, we integrate out the electronic degrees of freedom ($\Psi^{\dagger}$, $\Psi$) in order to quantify their influence in the dynamics of $\boldsymbol{\Delta}=(\Delta_S, \Delta_H)$. 

The integration of the electronic fields~\cite{altland2010condensed} yields
\begin{align}
Z=\int D(\Delta_S,\Delta_H)e^{-S_{\rm{eff}}(\Delta_S,\Delta_H)},
\end{align}
where $S_{\rm{eff}}=S_{\rm{ph}}-\Tr\ln G^{-1}$ is the effective action for the order parameters, and
\begin{equation}
    S_{\rm{ph}}=\frac{1}{2a} \int d\tau dx \left[m_{S}\left(\dot{\Delta}_S^{2}+\omega_{S}^{2}\Delta_S^{2}\right)+m_{H}\left(\dot{\Delta}_H^{2}+\omega_{H}^{2}\Delta_H^{2}\right)\right]\label{315}
\end{equation}
is the phonon-only part of the effective action.
In addition, $G^{-1}=\partial_{\tau}-\mu+h+\delta h$ is the inverse of the electron Green's function, and $\Tr$ represents the trace over both space-time and the $(L,R)$ pseudospin. 

From the effective action, we obtain the dynamics of the order parameters in two steps: first, we minimize the effective action via $\delta S_{\rm{eff}}/ \delta\boldsymbol{\Delta}=0$, and afterwards we switch back to real time. There is a problem, however. While the expression for $S_{\rm{eff}}$ is exact in principle, in practice it cannot be computed exactly because $G^{-1}$ contains order parameter fields with arbitrary time and space dependence. As a result, we are unable to compute the trace over space and time, unless we resort to approximations.

The simplest approximation is to assume that the external fields are weak and that the deviations of the order parameter $\boldsymbol{\Delta}$ from its mean-field value are small.
This approximation is unsatisfactory for the purposes of studying the switching of the electronic band topology, where $\boldsymbol{\Delta}$ must be allowed to flip by 180 degrees from its initial state.
A better approximation is to allow the direction of $\boldsymbol{\Delta}$ to vary arbitrarily, while assuming that the variation is slow in space and time.
To implement the latter approximation, it is convenient to carry out a unitary transformation known as the chiral rotation.

\subsection{Chiral rotation}

Let us define
\begin{align}
   -4\alpha_{S}\Delta_S&\equiv\Delta\sin\phi \notag
\\ \alpha_{H}\Delta_H&\equiv\Delta\cos\phi,\label{newangle}
\end{align}
where $\Delta$ is the modulus of the combined SSH and Holstein order parameters (in energy units), and $\phi$ is an angular variable characterizing the relative weight between the SSH and Holstein order parameters. 
Then, the electronic part of the action in Eq.~(\ref{39}) is
\begin{equation}
S_{\rm{e}}=\int dxd\tau \Psi^{\dagger}\left[\partial_{\tau}-iv\sigma^{z}\partial_{x}-\Delta\mathbf{\Omega}\cdot\boldsymbol{\sigma}-\mu+\delta h\right]\Psi,
\end{equation}
where  $\boldsymbol{\Omega}=(\cos\phi, \sin\phi,0)$.
Now consider a change of variables $\Psi'=R\Psi$, where $R=\exp\left(-i\sigma_{z}\phi/2\right)$ is a unitary operator that rotates the left- and right-moving fermions in the opposite directions by an angle $\phi$. 
A similar transformation has been used to study the dynamics of incommensurate charge density-waves~\cite{su1988inhomogeneity,yakovenko1998influence,krive1985evidence} and ferromagnets~\cite{spintronics}.
Upon this transformation, the electronic action becomes 
\begin{equation}
\label{eq:SD}
S_{\rm{e}}=\int dxd\tau\bar{\Psi}'\left(\slashed{D}'+M\right)\Psi',
\end{equation}
where $\bar{\Psi}'=\Psi '^{\dagger}\sigma^{x}$, 
$\slashed{D}'=\gamma^{\mu}\partial_{\mu}-ie\gamma^{\mu} A'_{\mu}$, $\gamma^{0}=\sigma^{x}$, $\gamma^{1}=-\sigma^{y}$, $\partial_0=\partial_\tau$, $\partial_1=v\partial_x$,
$A_0=iB_0/e$ and  $A_1=-B_z/e$. We define
\begin{align}
\label{eq:gauge_fields}
A'_0&=A_0-iv\partial_{x}\phi/2e\notag\\
A'_1 &=A_1+i \partial_{\tau}\phi/2e
\end{align}
as the effective electromagnetic potentials including the contributions from the gradients of $\phi$ \cite{yakovenko1998influence}.
The mass term in Eq.~(\ref{eq:SD})  is 
\begin{equation}
\label{eq:M}
M=\Delta+\mathbf{B}\cdot\mathbf{\Omega}-i\sigma^{z}(\mathbf{B}\times\mathbf{\Omega})\cdot\hat{\mathbf{z}}.
\end{equation}
In Eq.~(\ref{eq:SD}), the time and space gradients of $\phi$ have been separated out from $\phi$ itself. The later only appears in terms that multiply ${\bf B}$, which can be treated perturbatively. Thus, this opens the door of doing perturbation theory in $\partial_{\tau}\phi$ and $\partial_{x}\phi$, instead of $\phi$ itself.

\subsection{Chiral anomaly}

In Eq.~(\ref{eq:SD}), we have seen how the electronic action changes under a chiral rotation, through $(A_0,A_1)\to (A'_0, A'_1)$. As it turns out, when such an operation is applied to Dirac fermions in even space-time dimensions, the measure of the path integral changes as well~\cite{fujikawa1979path, roskies1981comment, fujikawa2004path}. This change results in an additional term $S_a$ to the effective action, 
\begin{equation}
\label{eq:Za}
Z =\int D[\Psi'^\dagger,\Psi',\phi,\Delta] e^{-S_a} e^{-S[\Psi'^\dagger,\Psi',\phi,\Delta]},
\end{equation}
where
\begin{equation}
S_{a}=\frac{1}{2\pi}\int d\tau dx \left[ie\phi E_{x}+\frac{1}{4 v}\left(\partial_{\tau}\phi\right)^{2}+\frac{v}{4}\left(\partial_{x}\phi\right)^{2}\right]
\label{eq:canom}
\end{equation}
is known as the chiral anomaly contribution and $E_x=-\partial_0 A^1-\partial_1 A^0$ is the electric field along the chain (not containing the contributions from $\partial_t\phi$ and $\partial_x\phi$ in Eq.~(\ref{eq:gauge_fields})).
Such chiral anomaly contribution to the total action in incommensurate CDW systems has been considered in   Refs.~[\onlinecite{su1988inhomogeneity,yakovenko1998influence,krive1985evidence}].

\subsection{Modulus of the order parameter and anisotropy energy}

In Eq.~(\ref{eq:Za}), there remains a difficulty when integrating out the fermion fields ($\Psi',\bar{\Psi}')$.
Namely, the resulting term 
${\rm Tr}\ln \left(\slashed{D}'+M\right)$ cannot be calculated explicitly because $\Delta$ in Eq.~(\ref{eq:M}) is generally space- and time-dependent. 
To make progress, we draw an analogy with an approximation that is commonly invoked  in magnetism~\cite{spintronics}. 

For temperatures well below the Curie temperature, the fluctuations in the magnitude
of the magnetization of a ferromagnet are relatively unimportant, because they involve higher energies than the fluctuations in the direction of the magnetization. In our model, the magnitude of the magnetization maps to   $\Delta$, while its direction maps to $\phi$. Then, if we assume that temperature is low compared to the mass gap,  $\Delta$ can be obtained by minimizing the mean-field energy $E[\Delta,\phi]$ with respect to $\Delta$, for each value of $\phi$. This amounts to neglecting the fluctuations of $\Delta$ around its mean-field value. 

Ignoring the contributions from $\mathbf{B}$ (whose effect on the dynamics of $\phi$ will be taken into account below), the energy minimization in the continuum model gives
 \begin{equation}
\Delta\simeq 4 at_{0}\Lambda\exp\left(-\lambda_S^{-1}\sin^2\phi -\lambda_H^{-1}\cos^2\phi \right).
 \end{equation}
Thus, the value of $\Delta$ becomes tied to that of $\phi$.
Assuming $\lambda_S \simeq \lambda_H$, we have 
\begin{equation}
\label{eq:Delta}
\Delta\simeq \Delta_{0}(1-\xi\sin^2\phi) \equiv\Delta(\phi),
\end{equation}
where the constant $\Delta_0$ 
is the value of $\Delta$ for $\lambda_S = \lambda_H$, and
\begin{equation}
\xi=\lambda_S^{-1}-\lambda_H^{-1}
\end{equation}
is a small dimensionless number.
Hence, we have succeeded in separating $\Delta$ into a dominant constant piece $\Delta_0$ plus a small time- and space-dependent piece.
The latter can be treated perturbatively when computing ${\rm Tr}\ln \left(\slashed{D}'+M\right)$.

Substituting Eq.~(\ref{eq:Delta}) in $E[\Delta,\phi]$, we get 
\begin{equation}
\label{eq:Ephi2}
E[\Delta,\phi]\simeq  {\rm const} + L \gamma \sin^2\phi \equiv E[\phi],
\end{equation}
where $L$ is the system length and
\begin{equation}
\gamma=\frac{\Delta_0^2 \xi}{4\pi a t_0} 
\end{equation}
is the analogue of the magnetic anisotropy energy per unit length for a two-dimensional magnet with an in-plane easy axis.
Like in the magnetic case, the anisotropy energy in the SSH-Holstein model at $\lambda_S\simeq \lambda_H$ is small compared to the magnitude of the order parameter.
When $\lambda_H>\lambda_S$, $E[\phi]$ is minimized for $\phi=0,\pi$ (degenerate minima). When $\lambda_H<\lambda_S$, $E[\phi]$ is minimized for $\phi=\pi/2, 3\pi/2$ (degenerate minima).
This is consistent with Fig.~\ref{fig:energy}. 
As mentioned above, for the purposes of the topological bit we choose $\lambda_H<\lambda_S$, so that in the absence of external perturbations the stable minima are $\phi=\pi/2$ and $\phi=3\pi/2$ (mod $2\pi$).

\subsection{Perturbative contributions to the effective action}

We are now ready to obtain an explicit expression for the effective action. 
Replacing $\Delta$ by $\Delta(\phi)$ in Eq.~(\ref{eq:Za}) and integrating out the fermion fields ($\Psi',\bar{\Psi}')$, we have
\begin{equation}
Z \simeq \int D[\phi] e^{-S_{\rm eff}[\phi]},
\end{equation}
where $S_{\rm eff} = S_{\rm ph} + S_a - {\rm Tr}\ln G'^{-1}$ and 
   $ G'^{-1} = \slashed{D}'+M$ is the inverse of the electronic Green's function after chiral rotation.
   Now, the effective action is only a function of $\phi$.
   The equation of motion for the order parameter follows from $\delta S_{\rm eff}/\delta\phi =0$.

In order to further develop the last term in $S_{\rm eff}$, we define the unperturbed inverse electronic Green function as
\begin{equation}
G_{0}^{-1}=\gamma^\mu\partial_\mu +\Delta_0
\end{equation}
and a perturbation $V=G^{-1}_{0}-G'^{-1}$ that is small 
provided that $(\partial_{t}\phi)/\Delta_0$, $v(\partial_{x}\phi)/\Delta_0$, $\xi$ and $|\mathbf{B}|/\Delta_0$ are small parameters. 
Yet, there is no restriction on the value of $\phi$.
Thus, a perturbative treatment on $V$ does allow to explore arbitrary values of $\phi$; such is the gain from the chiral rotation carried our in the preceding subsection. 

Expanding ${\rm Tr}\ln G'^{-1}$ to second order in $V$, the effective action reads
\begin{equation}
S_{\rm eff}[\phi]=S_{\rm{ph}}[\phi]+S_a+ \Tr(G_{0}V)+\frac{1}{2}\Tr(G_{0}VG_{0}V),
\label{eq:seff}
\end{equation}
where we have omitted a term that is independent of $\phi$ and thus does not contribute to the equation of motion of the order parameter.

The first order term in the perturbation gives
\begin{align}
{\rm Tr}(G_{0}V)&=-\frac{\Delta_{0}}{2\pi t_0 a \lambda_H}\int dxd\tau\Big(B_{x}\cos\phi\notag
\\
&+B_{y}\sin\phi-\Delta_0\xi\sin^2\phi\Big)~,\label{432}
\end{align}
where we have assumed that $\int dx B_0(x) =0$ (or, alternatively, we have absorbed the zeroth Fourier mode of $B_0$ into the chemical potential and have imposed that the renormalized $\mu$ stays in the middle of the mass gap).
Equation~(\ref{432}) is obtained in the standard way \cite{nagaosa1999quantum} by writing the trace in frequency and wave vector space, performing the Matsubara sum, taking the zero-temperature limit and finally computing the integral over the wave vector. We have also used that $v\Lambda\gg \Delta_0$.

The second order term in the perturbation, ${\rm Tr}(G_{0}V G_0 V)$, can be computed similarly but is singular. First, the result depends on the order in which the momentum and frequency integrals are carried out. Second, the result leads to an action that breaks gauge-invariance. Both of these problems can be remedied through a Pauli-Villars regularization. Upon regularizing, we find that the second order term in $V$ makes a negligible contribution to the effective action in comparison to $S_a$. More details about this point can be found in the Appendix.

\subsection{Effective action}

Combining the different terms of Eq.~(\ref{eq:seff}), the effective action for $\phi$ in real time reads
\begin{align}
S_{\rm eff}[\phi] \simeq 
&\int dxdt\left[\frac{(\partial_{t}\phi)^{2}-(v\partial_{x}\phi)^{2}}{8\pi v}+\gamma \sin^2\phi-\frac{e\phi E_{x}}{2\pi}\right.\notag\\
&-\frac{\Delta_{0}}{\pi v \lambda_H}\left(B_{x}\cos\phi +B_{y}\sin\phi\right)\notag\\
& \left.+\frac{\Delta^{2}_{0} m_H}{2a\alpha_H^2}\left(\frac{\omega_H^2}{\omega_S^2}\cos^{2}\phi+\sin^{2}\phi\right)(\partial_{t}\phi)^{2}\right],\label{530}
\end{align}
where the last line comes from the kinetic part of $S_{\rm ph}$ (the potential part of $S_{\rm ph}$ contributes to the anisotropy term $\gamma \sin^2\phi$). In the derivation of Eq.~(\ref{530}), we have assumed $\lambda_S\simeq \lambda_H$ and neglected higher order terms like $(\lambda_H-\lambda_S) (\partial_{t}\phi)^{2}$ and $(\lambda_H -\lambda_S)^2$.

\begin{figure}
\includegraphics[width=\hsize]{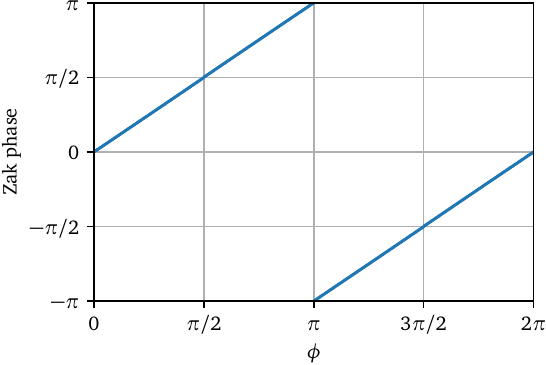}
\caption{Electronic Zak phase of the occupied band in the SSH-Holstein model, as a function of the order parameter angle $\phi$, for $\Delta\simeq 0.1 t_0$. The Zak phase matches with $\phi$ (modulo $2\pi$), thereby supporting Eq.~(\ref{eq:zak}) of the main text. \label{fig:zak}}
\end{figure}

Let us discuss two relevant aspects of $S_{\rm eff}$.
First, the electric polarization is given by
\begin{equation}
\label{eq:pol}
P = \frac{\delta S_{\rm eff}}{\delta E_x} = \frac{e\phi}{2\pi}.
\end{equation}
According to the modern theory of polarization \cite{vanderbilt2018berry}, when $\phi$ is constant or slowly varying we should have
\begin{equation}
\label{eq:zak}
\phi =i\int_{-\pi/a}^{\pi/a} dk\,  \langle u_k| \partial_k |u_k\rangle \,\,\,\,\,\,\,\text{      mod $2\pi$},
\end{equation}
where $|u_k\rangle$ is the eigenstate of the occupied band in the so-called canonical electronic Bloch Hamiltonian \cite{cayssol2021topological}.
Figure~\ref{fig:zak} displays the right hand side of Eq.~(\ref{eq:zak}) as a function of $\phi$, and confirms that Eq.~(\ref{eq:zak}) is indeed satisfied for $\Delta_0/t_0\ll 1$. 
The departures from Eq.~(\ref{eq:zak}) become noticeable when $\Delta_0/t_0\gtrsim 1$. This can be attributed to the fact that the chiral anomaly action responsible for Eq.~(\ref{eq:pol}) is calculated from the low-energy continuum model, which breaks down for $\Delta_0\gtrsim t_0$, whereas the electric polarization is computed from the full lattice model.

We now turn to a second aspect of $S_{\rm eff}$. The  electric charge ($\rho$) and current ($j$) densities are given by
\begin{align}
\label{j0}
\rho&=\frac{\delta S_{\rm eff}}{\delta A^0}= -\frac{e}{2\pi}\partial_{x}\phi\notag\\
j&=-\frac{\delta S_{\rm eff}}{\delta A^1}= \frac{e}{2\pi}\partial_{t}\phi.
\end{align}
At a sharp domain wall between $\phi=\pi/2$ and $\phi=-\pi/2$ centered at $x=0$, Eq.~(\ref{j0}) implies $\rho(x)=-  e\delta(x)/2$. Thus, a particle of charge $e/2$ is trapped at the domain wall between two regions of different topological invariant. This result is well-known in the SSH model literature~\cite{rajaraman1982solitons, asboth2016short}.
It is less frequently mentioned \cite{cayssol2021topological} that, as evidenced by Eq.~(\ref{j0}), a particle of charge $e/2$ can also be trapped at any domain wall where $\phi$ jumps by $\pi$.  

Hereafter we focus on spatially uniform order parameter dynamics ($\partial_x\phi=0$) in the bulk of the chain, neglecting the effect of chain boundaries
\footnote{In practice, we must consider finite-sized chains. Otherwise, there would be no way to physically distinguish $\phi=\pi/2$ from $\phi=-\pi/2$.
Yet, for technical simplicity, we take the approach of treating the dynamics of the order parameter as though the chain were infinitely long and uniform.
This approach is approximately valid for long enough chains, where the effect of the boundaries on the equation of motion of the order parameter is relatively unimportant.}.  
This is analogous to the case of magnetic memories based on single-domain ferromagnets.

\subsection{Inclusion of damping}

One key ingredient of magnetic memories that is missing so far in the topological counterpart is damping. 
Energy loss is essential in order to ensure that the order parameter departing from, say, $\phi=-\pi/2$ under the action of a transient external perturbation will eventually be able to relax into a local energy minimum corresponding to $\phi=\pi/2$.

Damping of the order parameter in the SSH-Holstein model originates from electron-phonon coupling and phonon-phonon coupling. The electronic mechanism for damping is the decay of order parameter fluctuations into particle-hole pairs. This is analogous to the decay of magnons into electron-hole pairs in conducting ferromagnets.
However, this process is suppressed at zero temperature, when the chemical potential is inside the energy gap and when the characteristic frequency of the dynamics of the order parameter is smaller than the energy gap (which we have assumed above).
We thus concentrate on the intrinsic phonon damping, which is analogous to magnon-magnon coupling in magnetic relaxation.

We incorporate such damping phenomenologically through the Caldeira-Leggett model~\cite{caldeira1983path}.
We suppose that SSH and Holstein phonons are coupled to baths of harmonic oscillators. 
This results in an additional term that must be added to the real-time action (for a succinct review, see e.g.  Ref.~[\onlinecite{wen2004quantum}]),
\begin{align}
S_{\rm damp} &= \frac{1}{4}\int dx dt dt'  \left\{G_S(t-t') \left[\Delta_S(x,t)-\Delta_S(x,t')\right]^2\right.\notag\\
&\left. +G_H(t-t') \left[\Delta_H(x,t)-\Delta_H(x,t')\right]^2\right\},
\end{align}
where $G_i(t)=G_i(-t)$ is the time-ordered correlation function of the bath for SSH ($i=S$) and Holstein ($i=H$) phonons.
For simplicity, we have assumed that the damping is local in space.

The contribution of $S_{\rm damp}$ to the equation of motion renormalizes the oscillator frequencies and provides a friction term, provided that $G_i(t)$ is replaced (ad hoc) by the retarded correlation function \cite{wen2004quantum}.
Herein, we neglect the frequency renormalization and concentrate solely on the friction term.
Then, the contribution of $S_{\rm damp}$ to the equation of motion for $\phi$ is
\begin{equation}
\label{eq:dSdp}
\frac{\delta S_{\rm damp}}{\delta \phi}= \frac{\delta S_{\rm damp}}{\delta \Delta_S}  \frac{\partial\Delta_S}{\partial \phi} +  \frac{\delta S_{\rm damp}}{\delta \Delta_H}   \frac{\partial\Delta_H}{\partial \phi},
\end{equation}
where
\begin{equation}
\label{eq:dSdD}
\frac{\delta S_{\rm damp}}{\delta \Delta_S} = \frac{m_S}{a}\Gamma_S \partial_t \Delta_S\text{  ;  } 
\frac{\delta S_{\rm damp}}{\delta \Delta_H}  = \frac{m_H}{a}\Gamma_H \partial_t \Delta_H,
\end{equation}
$\Gamma_i \propto {\rm Im}[G_i(\omega)/|\omega|]$ is the damping rate and $G_i(\omega)$ is the Fourier transform of $G_i(t)$ at frequency $\omega$. 
In the Ohmic damping approximation that we adopt, $\Gamma_i$ is  a constant.
Hereafter, we will assume for simplicity that $\Gamma_S=\Gamma_H \equiv \Gamma$.
Then, Eq.~(\ref{eq:dSdD}), combined with the bare phonon action $S_{\rm ph}$, results in damped harmonic oscillations with a damping rate $\Gamma$ for both SSH and Holstein order parameters.

Neglecting smaller terms like $(\lambda_H^{-1}-\lambda_S^{-1})\Gamma/t_0$, we have
\begin{equation}
\label{eq:damping}
\frac{\delta S_{\rm damp}}{\delta \phi}
\simeq \frac{\Delta_0^2 m_H }{a \alpha_H^2}\left(\frac{\omega_H^2}{\omega_S^2} \cos^2\phi + \sin^2\phi\right) \Gamma \partial_t \phi,
\end{equation}
which is expectedly invariant under $\phi \to \phi + 2\pi$.

\subsection{Equation of motion for the order parameter}
\label{sec:mot}
Starting from Eq.~(\ref{530}), taking into account the damping contribution (Eq.~(\ref{eq:damping})), and restoring all factors of $\hbar$, the condition $\delta(S_{\rm eff}+S_{\rm damp})/\delta\phi=0$ results in
\begin{align}
&(1+\eta)\partial_{t}^{2}\phi +\eta\Gamma\partial_{t}\phi+\frac{4\pi v\gamma}{\hbar}\sin2\phi\notag \\
&=\frac{\eta\omega_{H}^{2}}{\Delta_0}\left(-B_x \sin\phi + B_y \cos\phi\right)-\frac{4e v E_{x}}{\hbar},\label{4588}
\end{align}
where 
\begin{equation}
\eta\equiv \frac{4\Delta_{0}^{2}}{\hbar^2 \omega_{H}^{2} \lambda_H} 
\end{equation}
 is a dimensionless parameter. In the regime of adiabatic phonons ($\hbar\omega_H\ll \Delta_0$) and continuum approximation ($\lambda_H\ll 1$), we expect $\eta\gg 1$.
In the derivation of Eq.~(\ref{4588}), we have assumed that $\omega_S=\omega_H$ \footnote{In real quasi one-dimensional crystals with intramolecular (Holstein) and intermolecular (SSH) phonons, the former typically have higher frequencies (remark courtesy of Prof. Claude Bourbonnais). Our assumption of $\omega_S=\omega_H$ is made on the grounds of analytical simplicity. Should this assumption be relaxed, the analogy with the dynamics of a Josephson junction (discussed in Sec.~\ref{sec:jj}) would no longer hold. Yet, the idea of the topological bit would still be realized}. 
Even though we are interested in the case $\gamma<0$ (where the ground state is an SSH bond density-wave), Eq.~(\ref{4588}) holds also for the case $\gamma>0$ (where the ground state is a Holstein site density-wave).
We also mention in passing that Eq.~(\ref{4588}) holds for neutral fermions as well, provided that $e=0$ is taken.

For numerical convenience, we can recast Eq.~(\ref{4588}) in dimensionless form.
First, we introduce a dimensionless time $\bar{t}=t/\tau_{a}$, 
where 
\begin{equation}
\tau_a=\sqrt{\frac{\hbar(1+\eta)}{4\pi v|\gamma|}}
\end{equation}
is the characteristic time associated to the anisotropy energy.
Then, Eq.~(\ref{4588}) becomes
\begin{align}
&\partial_{\bar{t}}^{2}\phi+\bar{\Gamma}\partial_{\bar{t}}\phi-\sin2\phi=-\bar B_{x}\sin\phi+\bar B_{y}  \cos\phi-\bar E_{x},\label{465}
\end{align}
where
\begin{align}
\label{eq:bars}
\bar{\Gamma}&=\frac{\eta}{1+\eta}\Gamma\tau_a\notag\\
\bar{B}_{x,y} &= \frac{\eta}{(1+\eta)} (\omega_H \tau_a)^2 \frac{B_{x,y}}{\Delta_0}\notag\\
\bar{E}_x &= -\frac{e}{\pi\gamma} E_x
\end{align}
are dimensionless parameters associated to damping, pseudomagnetic fields and electric field, respectively.

Several terms in Eq.~(\ref{4588}) are familiar from the literature on the sliding of incommensurate CDW~\cite{gruner1988dynamics}.
Such is the case for the inertial ($\partial_t^2\phi$), damping ($\partial_t\phi$) and electrical driving ($E_x$) terms. 
On the right hand side, the term proportional to $B_x$ has in the past been associated to the effect of impurities, which provide a $\cos\phi$ potential  that pins the incommensurate CDW.
One novelty in our theory is that the main pinning potential for the CDW originates instead from the anisotropy energy $\gamma$. 
This $\sin^2\phi$ pinning potential differs qualitatively from the one resulting from $B_x$ because it creates stable energy minima at $\phi=-\pi/2$ and $\phi=\pi/2$.
This property is crucial for the realization of the topological bit. 
Another novelty in Eq.~(\ref{4588}) is the presence of the term proportional to $B_y$.

Below, Eq.~(\ref{4588}) will allow us to establish the basis of the topological bit. But before, let us discuss the analogy between Eq.~(\ref{4588}) and the dynamics of a Josephson junction.

\section{Analogy with  an unconventional  Josephson junction}
\label{sec:jj}

The analogy between the CDW dynamics and the dynamics of a current-biased Josephson junction (JJ) is amply documented in the literature~\cite{gruner1988dynamics}.
In Eq.~(\ref{4588}), this analogy contains some novel elements that will be discussed in this section.

The current conservation equation in a Josephson junction can be approximately described by~\cite{likharev2022dynamics}  
\begin{equation}
\frac{\hbar C}{e}\partial_{t}^{2}\theta+\frac{\hbar}{eR}\partial_{t}\theta+I_{0}\sin2\theta= I_{\rm{bias}},
\label{461}
\end{equation}
where $I_{\rm{bias}}$ is the bias current, $2\theta$ is the superconducting phase difference across the junction, $R$ is the normal-state resistance of the weak link, and $C$ is the junction capacitance. The first term on the left-hand side of Eq.~(\ref{461}) is Maxwell's displacement current. The second term is the normal (quasiparticle) dissipative current, while the third term  is the dissipationless current due to the tunneling of Cooper pairs, with its maximum amplitude given by $I_0$. 
It is clear that, if we take $B_x=B_y=0$, Eq.~(\ref{461}) has the same form as Eq.~(\ref{4588}); see Table~\ref{tab:table} and Ref.~[\onlinecite{gruner1988dynamics}].

\begin{table}[h!]
  \begin{center}
    \caption{Correspondence between charge density-wave dynamics  (Eq.~(\ref{4588})) and Josephson junction dynamics (Eq.~(\ref{461})). The pseudomagnetic fields refer to $B_x$ and $B_y$.}       
         \label{tab:table}
    \begin{tabular}{ll}
    \hline\hline
	CDW & Josephson junction\\
          \hline
      	Order parameter angle ~~~ & Superconducting phase difference   \\
       	Anisotropy field & Conventional supercurrent \\
       	Pseudomagnetic fields & Unconventional supercurrents \\
         	Damping & Dissipative current \\
          Electric field & Current bias \\
          \hline\hline
    \end{tabular}
  \end{center}
\end{table}

  Taking advantage of the analogy with the Josephson effect, we can anticipate the effect of the electric field in the dynamics of $\phi$. To that end, we first recognize that (still for $B_x=B_y=0$) Eq.~(\ref{4588}) can be understood as the equation of motion of an effective particle moving in a tilted washboard potential 
\begin{equation}
\label{eq:wash}
U[\phi]=(2\pi v \gamma/\hbar) \cos2\phi-(4 e v E_{x}/\hbar)\phi,
\end{equation}
where $\phi$ plays the role of the position of the particle. If $e |E_{x}|<\pi|\gamma|$ (|$\bar{E}_x| <1$), there is a stable solution where $\phi$ makes small oscillations around a local energy minimum of $U(\phi)$. In this case, $E_x$ alone cannot induce a topological phase transition. 
In  Fig.~\ref{fig:energy}a, the system stays close to -- and makes small oscillations about -- a local energy minimum.

\begin{figure}
\includegraphics[width=\hsize]{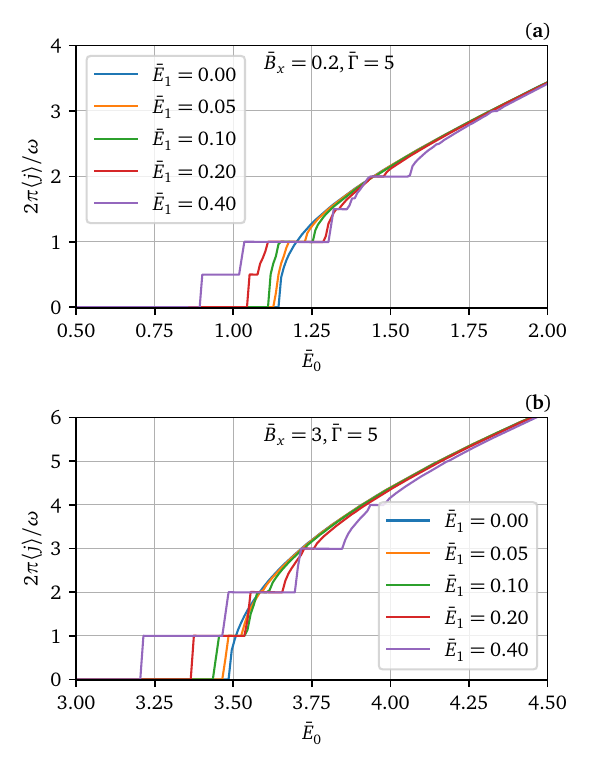} 
\caption{Shapiro steps in a voltage-biased SSH-Holstein chain. The time-averaged current $\langle j\rangle$ is calculated by solving Eq.~(\ref{465}) numerically in the presence of an electric field $E_x=E_0+E_1\cos(\omega t)$, where $E_0$ and $E_1$ are constants. A constant pseudomagnetic perturbation $B_x$ (which could describe e.g. the effect of static  impurities) is likewise included. The bars indicate dimensionless parameters (see Eq.~(\ref{eq:bars})).  (a) When $\bar{B}_x\ll 1$ (anisotropy dominates over pseudomagnetic field), Shapiro steps take place when $2 \pi \langle j\rangle/\omega$ is a half-integer. (b) When $\bar{B}_x\gg 1$ (pseudomagnetic field dominates over anisotropy),  Shapiro steps take place when $2 \pi \langle j\rangle/\omega$ is an integer. The doubling in the periodicity of Shapiro steps is analogous to the situation in a fractional Josephson junction.
\label{fj}}
\end{figure}
For $e |E_{x}|>\pi|\gamma|$ (|$\bar{E}_x| >1$), $U(\phi)$ becomes a monotonic function of $\phi$ and accordingly the effective particle keeps sliding down the washboard potential. 
In Fig.~\ref{fig:energy}a, the order parameter circles repeatedly along a rim in the energy landscape (white line). 
In real space, the charge density wave slides along the chain. 
In this case, while the electric field is on, the system undergoes a succession of topological phase transitions. Once the electric field is turned off, the system will once again relax towards a local minimum of the untilted washboard potential, aided by damping. If the final and the initial values of $\phi$ differ by $\pi$ (modulo $2\pi$), then we will have realized a permanent topological phase transition with a transient electric field. We will discuss this transition further in the next section.

Thus far, we have considered $B_x=B_y=0$. When $B_x\neq 0$, we get a new supercurrent-like term with a doubled periodicity in $\phi$: in Eq.~(\ref{4588}), $B_x$ multiplies $\sin\phi$ while $\gamma$ multiplies $\sin2\phi$. Therefore, a constant $B_x\neq 0$ gives rise to an effect that is analogous to the fractional Josephson effect. Recently, the fractional Josephson effect has attracted a lot of interest in Josephson junctions made out of topological superconductors~\cite{aguado2017majorana}. In these systems, Majorana bound states are responsible for the fractional Josephson effect. Intense efforts have been devoted to detect such an effect through unconventional Shapiro steps and Josephson radiation, for example. It is interesting that an analogue of the fractional Josephson effect could also take place in charge density-wave systems \footnote{A caveat about the semantics is in order. A static $B_x \cos\phi$ term has in the past been associated to the pinning potential for an incommensurate CDW, originating e.g. from impurities \cite{gruner1988dynamics}. Then, the term $B_x \sin\phi$ in the equation of motion has been referred to as the analogue of the conventional Josephson current. From that point of view, the anisotropy force term $\gamma\sin(2\phi)$ appearing in our theory would be a current of {\em pairs} of Cooper pairs. Yet, in the context of a commensurate CDW, it is natural to associate the $\gamma\sin(2\phi)$ term with the ordinary supercurrent and the $B_x \sin\phi$ term with a fractional supercurrent (dissipationless tunneling of single electrons), at least when the system is clean and $B_x$ is applied externally. At any rate, independently from the semantics, we may state that in the presence of $B_x$ and $\gamma$, the dynamics of a CDW is akin to that of an unconventional Josephson junction with two different coexisting periodicities in its supercurrent.}. 

The unconventional character of the analogue Josephson junction is further evidenced when $B_y\neq 0$. In this case, a term proportional to $B_y \cos\phi$ is added to the supercurrent. The coexistence of $\sin\phi$ and $\cos\phi$ terms in the supercurrent is known to occur in the so-called $\varphi_0$-junctions~\cite{buzdin2003periodic,szombati2016josephson}.

The existence of two different harmonics of $\phi$  in the equation of motion leads to unconventional Shapiro steps. Let us suppose a voltage-biased system with 
\begin{equation}
E_x=E_0+E_1\cos(\omega t),
\end{equation}
where $E_0$, $E_1$ and $\omega$ are constants.
Panels (a) and (b) of Fig.~\ref{fj} display the time-averaged electric current density $\langle j\rangle$, calculated numerically from Eq.~(\ref{465}), as a function of $E_0$ for the cases of strong ($\bar{B}_x\ll 1$) and weak ($\bar{B}_x\gg 1$) anisotropy. We set $B_y=0$ for simplicity.
If anisotropy is weak ($\gamma\to 0$), the Shapiro steps take place when  $\langle j\rangle$  is a multiple of $e \omega/(4\pi)$.
If anisotropy is strong ($B_x\to 0$), the Shapiro steps occur instead when $\langle j\rangle$  is a multiple of $e \omega/(2\pi)$, i.e. they are half as numerous.
These findings are in agreement with standard theory of Shapiro steps (see e.g. Ref.~[\onlinecite{tinkham2004introduction}] for the case of Josephson junctions).

Consequently, Shapiro steps provide a diagnostic tool for the phase transition between the SSH-like and Holstein-like CDW. 
At the phase boundary, the anisotropy energy $\gamma$ vanishes and the Shapiro steps arise when $\langle j\rangle =  n e \omega/(2\pi)$ with integer $n$. 
Sufficiently far from the phase boundary, $\gamma$ dominates over $B_x$ and hence the Shapiro steps occur when $\langle j\rangle =  n e \omega/(4\pi)$. 

\section{Switching Dynamics}
\label{sec:switch}

 \begin{figure}
\includegraphics[width=\hsize]{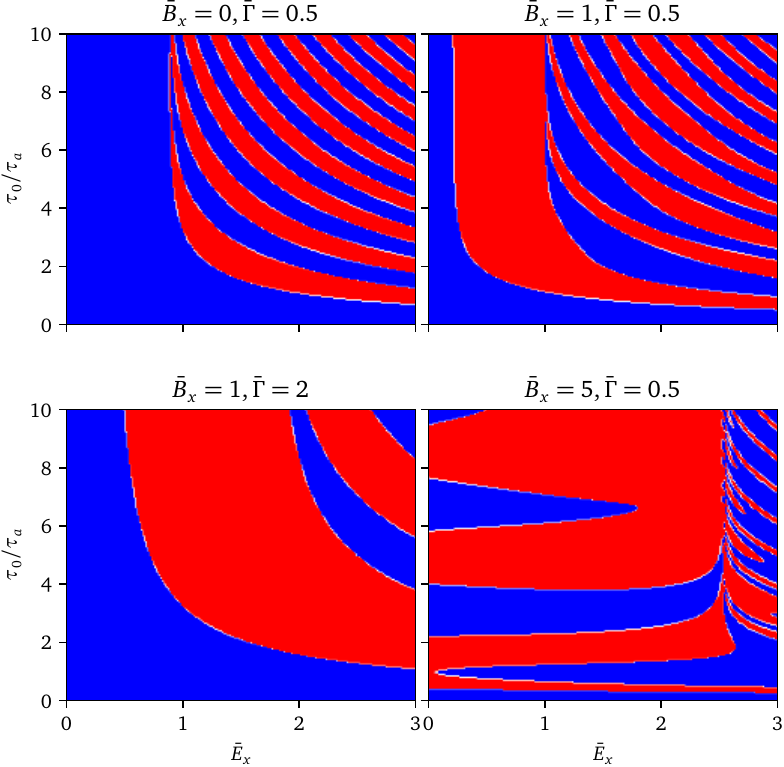}
\caption{Band topology of the system at long times, after being driven by external perturbations $\bar{E}_x$ and $\bar{B}_x$ of duration $\tau_0$. 
The parameters $\tau_a$ and $\bar{\Gamma}$ denote the anisotropy time and the damping strength. 
In the blue (red) regions, the topological invariant of the final state is the same as  (different from) the topological invariant of the initial state.\label{swd}}
 \end{figure}

In this section, we provide a proof of concept for the topological bit in the SSH-Holstein model. 
We take the initial state to be $\phi(0)=\pi/2\,\, {\rm mod}\pi$. 
This is a stable state in the absence of perturbations.
Then, we apply a square external pulse $(E_{x}(t) , B_{x}(t)) $ of finite duration $\tau_0$.
For simplicity, we keep $B_y=0$.
We also rule out a {\em static} $B_x$. 
Should such a static field be present (e.g. due to impurities), we would require that it be smaller than the anisotropy energy; otherwise, $\phi=\pm \pi/2$ would no longer be energy minima and the topological bit would not be realized.
 
Solving Eq.~(\ref{465}) numerically, we obtain the final state at long times, $\bar{t}\gg (\tau_0/\tau_a, 1/\bar{\Gamma})$.
The final state is of the form $\phi(\infty)= \phi(0)+n\pi$, where $n\in\mathbb{Z}$.
If $n$ is odd,  we will have achieved the desired topological switch.
The numerical results for $\phi(\infty)$  are displayed in the switching diagrams of Fig.~\ref{swd}. The blue (red) color corresponds to $n$ even (odd). Thus, the existence of red regions confirms the possibility of a topological bit.

When $B_x=0$,  there exists a critical value for $E_x$ beyond which the topological phase transition is realized. The critical $\bar{E}_x$ tends to $1$ as $\tau_0$ becomes long; this is consistent with the discussion below Eq.~(\ref{eq:wash}).
The switching diagrams contain alternating red and blue arcs, which are caused by the inertia of the CDW. Even after the pulse is turned off, the CDW will continue to slide until it is finally stopped by damping. 
When either $\tau_0$ or $E_x$ increases, the final state alternates between blue and red regions.  

For larger values of $B_x$, the red regions in the figures become wider overall (compare the top two panels of Fig. \ref{swd}). In other words, $B_x$ helps realizing the desired topological bit flip. To understand why that is the case, we recall that $B_x$ multiplies $\sin\phi$ in the equation of motion, whereas the anisotropy field scales as $\sin2\phi$. Thus, in the effective potential energy, $B_x$ multiplies $\cos\phi$, while the anisotropy energy goes like $\cos2\phi$. Plotting combined $\cos2\phi$ and $\cos\phi$ functions, we see that $B_x$ reduces the energy barrier for transitions between topologically distinct states (e.g. $\phi \to \phi+\pi$), while it increases the energy barrier for transitions between topologically identical transitions (e.g $\phi\to\phi+2\pi$).  

In all the graphs there exists a critical value of $\tau_0$ beyond which the topological phase transition is realized.
Obviously, if $\tau_0$ is very short, finite external perturbations cannot contribute significantly to the equation of motion~(\ref{465}).
The values of both $B_x$ and $E_x$ have significant impacts on the critical value of $\tau_0$.  When $E_{x}$ is large, a small increase of $\tau_0/\tau_a$ leads to a large change of $\phi$ in the final state. This is why successive blue and red regions get closer at stronger electric fields. As a result, for strong $E_x$, it is harder to predict whether the final value of $\phi$ will be the correct one (red) or the wrong one (blue). If $E_{x}$ becomes infinite, the final value of $\phi$ becomes unpredictable. 
The most favorable combination for a roust and controllable topological switch appears to be $\bar{E}_x\simeq \bar{B}_x\lesssim 1$.

The results in Fig. \ref{swd} were obtained for moderately strong damping. With stronger damping, the width of the arcs gets generally larger, thereby facilitating a controlled switch of the topological invariant (compare the top right and bottom left panels of Fig.~\ref{swd}). To the contrary, if damping is weak, the system behaves closer to the undamped case and $\phi(\infty)$ becomes very sensitive to small changes in $\tau_0$, $E_x$ and $B_x$.


\section{Discussion and conclusion}
\label{sec:conc}

In summary,  we have constructed a proof-of-principle for the topological analogue of a magnetic bit, where a relatively small and transient external perturbation can lead to a permanent change in the electronic band topology.
We have theoretically illustrated this analogy in the SSH-Holstein model, where two distinct charge density-wave order parameters (the SSH-like $\Delta_S$ and Holstein-like $\Delta_H$) compete with one another.  We have considered the situation in which the SSH-like order parameter is realized in the ground state, with a uniform value of the order parameter. 
 This is  akin to a single-domain ferromagnetic order.

Then, using a time-dependent mean-field theory, we have obtained the equation of motion for the order parameter and have confirmed that it can continuously rotate by $180$ degrees in the $(\Delta_S, \Delta_H)$ plane, under an appropriate external perturbation. 
This process is analogous to a magnetic bit flip.
In real space, the reversal of the charge density wave order parameter is tantamount to a displacement of the electronic system by an odd number of unit cells. 
The resulting change in the momentum-space winding number of the electronic wave functions has observable effects on the boundaries of a finite-size chain. 

When the number of sites in the chain is even, the logical 0 can be ascribed to the ground state without edge modes, while the 1 state has two topologically protected edge modes (one on each boundary).
When the number of sites is odd, the logical 0 can be associated to the state that has an edge mode on the right boundary, whereas the 1 state has an edge mode on the left boundary
\cite{zaimi2021detecting}.
Irrespective of having an even or an odd number of sites, we may attribute the logical 1 (0) to the ground state in which the left boundary has (does not have) an edge mode. 
In all cases, the states 0 and 1 differ by a change in sign of $\Delta_S$.

To read such a bit, one should be able to probe the local density of states on the boundary of a chain.
To our knowledge, the edge states of the SSH model have not been directly seen in condensed matter settings. 
Recent implementations of the SSH model have taken place in the context of photonic crystals \cite{st2017lasing, caceres2022experimental}, acoustic crystals \cite{li2018schrieffer}, and cold atomic systems \cite{atala2013direct, meier2016observation,kanungo2022realizing}. 
In these systems, local probes have been able to detect edge modes.
Yet, the SSH-Holstein model where $\Delta_S$ and $\Delta_H$ are dynamical {\em order parameters} (as opposed to externally set parameters) remains to be experimentally realized.

The equation of motion for the order parameter that we have obtained resembles the one corresponding to incommensurate charge density-wave systems, derived e.g. in
~\cite{gruner1988dynamics,eckern1986microscopic,yakovenko1998influence,su1988inhomogeneity,krive1985evidence,ishikawa1988phase}. 
Comparing with those earlier papers, the main novelties in our work are that (i) we consider a commensurate charge density-wave with nonzero anisotropy energy, while preserving the energy degeneracy between two topologically distinct ground states; (ii) we consider the effect of external perturbations other than an electric field; (iii) we seek to realize a proof of principle for a topological bit, by identifying situations where the charge density-wave slides by an odd number of unit cells. Points (i) and (ii) are essential for point (iii); as such, our work is distinct from the earlier literature on the dynamics of charge density-waves.

Most recently, the possibility of producing a topological phase transition in a one-dimensional charge density-wave system with a time-dependent external perturbation has been theoretically explored \cite{zhang2021ultrafast}.
While relevant, this numerical study is qualitatively different from our work in a number of ways. First, it treats the dynamics of an incommensurate charge density-wave. Second, it considers a light pulse that induces interband electronic transitions across the insulating energy gap (contrary to our case, where the dynamics is slower than the energy gap). Third, the light-pulse in Ref.~\cite{zhang2021ultrafast} is spatially inhomogeneous (we work with uniform external perturbations). Fourth, the light-induced topological phase predicted in Ref.~\cite{zhang2021ultrafast} consists of the winding of the CDW order parameter phase in real space (as opposed to momentum space, like in our case).

Looking forward, there are several research directions that deserve further study. For instance, it would be interesting to generalize the topological analogue of the magnetic bit to higher dimensional systems. 
Moreover, it remains an open question whether one can identify an experimentally realizable platform where the analogy can be implemented. 
Candidate systems include quasi-one dimensional crystals with intramolecular and intermolecular phonons, or quasi one-dimensional spin systems with phonon-modulated spin-spin interactions.

\acknowledgements
This work has been financially supported by the Natural Sciences and Engineering Research Council of Canada (Grants No. RGPIN- 2018-05385 and 2020-05060), and the Fonds de Recherche du Qu\'ebec Nature et Technologies.
We acknowledge very informative discussions with C. Bourbonnais, as well as feedback from A. Blais and R. Côté.
I. G. is grateful to P. L. S. Lopes for valuable contributions in the initial stages of the project, as well as feedback on the manuscript. 


\appendix

\section{Details about the calculation of $\Tr\left(G_{0}VG_{0}V\right)$}

In the main text, we have stated that the second order perturbative term in Eq.~(\ref{eq:seff}) does not make a substantial contribution to the effective action. 
In this appendix, we substantiate that assertion.

The second order term $\Tr\left(G_{0}VG_{0}V\right)$ can be written in Fourier space as
\begin{equation}
\label{eq:bub}
\Tr\left(G_{0}VG_{0}V\right)\propto \frac{1}{L \beta}\sum_{k, q} \tr\left[G_0(k) V(q) G_0(k-q) V(-q)\right],
\end{equation}
where $\beta$ is the inverse temperature, $\tr$ denotes the trace over the pseudospin, and $k$ (or $q$) denotes both wave vector and frequency.
The next step is to do the $k$-sum in order to get an explicit contribution to the effective action.
In the condensed matter literature, it is customary to carry out the frequency (Matsubara) sum first, followed by the sum over the wave vector. 
Doing so and expanding the result to leading order in $q$, we get
\begin{equation}
\label{4655}
{\rm Tr}\left(G_{0}VG_{0}V\right) = -\frac{e^2}{4\pi v} \sum_q A'_1(q) A'_1(-q) +\cdots~,
\end{equation}
where we have taken the zero-temperature limit after the Matsubara summation, and we have omitted terms involving $B_x$, $B_y$ and $\xi$ (as those already appear at first order in perturbation theory).
In addition, we have kept only the lowest-order contribution in space and time gradients.
At first glance, Eq.~(\ref{4655}) is problematic as it is not gauge-invariant (cf. Eq.~(\ref{eq:gauge_fields})). 

If we reverse the order between frequency and wave vector sums in Eq.~(\ref{eq:bub}), we still get the same result if the wave vector sum has a finite cutoff $\Lambda$ and we take $\Lambda\to\infty$ in the very end. 
But, if we insist on doing the wave vector sum first from $-\infty$ to $\infty$ and then the Matsubara sum, we get (under the same approximations as above)
\begin{equation}
\label{434}
{\rm Tr}\left(G_{0}VG_{0}V\right) = \frac{e^2}{4\pi v} \sum_q A'_0 (q) A'_0 (-q)+\cdots~,
\end{equation}
which differs from Eq.~(\ref{4655}) and is not gauge-invariant either.

To find the origin of the preceding problems, we notice that, in the zero-temperature limit, ${\rm Tr}\left(G_{0}VG_{0}V\right)$ involves the following integrals:  
\begin{equation}
\label{eq:intill}
\int^{\infty}_{-\infty} dk_{0}dk_{1} \frac{(\mp k_{0}^{2}\pm v^{2}k_{1}^{2}+\Delta_{0}^{2})}{(k_{0}^{2}+v^{2}k_{1}^{2}+\Delta_{0}^{2})^2}, 
 \end{equation}
whose outcome varies depending on whether we do the frequency ($k_0$) or the wave vector ($k_1$) integral first.
In polar coordinates, where $k_0=r \cos\theta$ and $k_1=(r/v) \sin\theta$, Eq.~(\ref{eq:intill}) reads
\begin{equation}
\label{4400}
\int^{2\pi}_{0}d\theta\int^{\infty}_{0}dr\frac{1}{v}r\frac{\Delta_{0}^{2}\mp r^{2}\cos2\theta}{(r^{2}+\Delta_{0}^{2})^{2}}.
\end{equation}
 This integral is ill-defined, in the sense that  
\begin{equation}
\int^{2\pi}_{0} d\theta\left(\int^{\infty}_{0} dr\frac{r^{3}  \cos2\theta}{(r^{2}+\Delta_{0}^{2})^{2}}\right)=\infty
\end{equation}
and 
\begin{equation}
\int_0^\infty dr\left(\int^{2\pi}_{0} d\theta\frac{r^{3}  \cos2\theta}{(r^{2}+\Delta_{0}^{2})^{2}}\right)=0~.
\end{equation}
One recipe to solve this problem is to apply a UV regularization, such as the Pauli-Villars regularization~\cite{peskin2018introduction}, which consists in subtracting from ${\rm Tr}(G_0VG_0V)$ the same quantity but with the  Dirac mass $\Delta_0$ replaced by another mass $M$. In the end of the calculation, $M$ is taken to infinity.
Because neither Eq.~(\ref{4655}) nor Eq.~(\ref{434}) depends on the mass to leading order in $q$, the regularized ${\rm Tr}(G_0VG_0V)$ vanishes to leading order in $q$ irrespective of the order in which the frequency and momentum integral is carried out.
Gauge noninvariant terms are therefore eliminated.
This does not mean that the regularized ${\rm Tr}(G_0VG_0V)$ is exactly zero: it has higher-order derivative terms in $\phi$ and in the gauge fields, but those terms are gauge-invariant and small compared to the ones that appear in the action terms we have kept in the main text. They can therefore be neglected. 

We conclude this appendix by comparing our approach to a number of earlier studies, which, in the course of investigating the dynamics of incommensurate charge density-waves ~\cite{eckern1986microscopic,yakovenko1998influence,su1988inhomogeneity,krive1985evidence,ishikawa1988phase}, discussed in some form the calculation of $ {\rm Tr}(G_{0}VG_0V)$.
While all of these works reached similar results for the equation of motion describing the dynamics of the CDW, their procedures and arguments differ substantially from one another and can altogether give rise to confusion. 
It must also be mentioned that those works, which we discuss briefly next, omit the external perturbations ($B_x, B_y$) and the anisotropy energy $\gamma$ (the CDW treated therein is incommensurate).

In Ref.~\cite{eckern1986microscopic}, the authors assume that all the electronic contributions to the effective action of the order parameter come from ${\rm Tr}(G_0VG_0V)$. There is no discussion of the change in the path integral measure under the chiral rotation, and hence the chiral anomaly action is apparently overlooked. In addition, the authors' choice of a concrete gauge hides the fact that the calculated ${\rm Tr}(G_0VG_0V)$ is not gauge-invariant.

Reference~\cite{yakovenko1998influence} adopts a semi-relativistic model for fermions: while a linear electronic dispersion is used, a diamagnetic term of the electromagnetic vector potential is also included (unlike in our relativistic model). This diamagnetic term contributes to ${\rm Tr}(G_0VG_0V)$ and cancels a term that is not gauge-invariant.
The resulting expression for ${\rm Tr}(G_0VG_0V)$ still breaks gauge invariance, but the authors manage to restore it through the computation of the chiral anomaly term. 
Unlike in our theory, the chiral anomaly term in Ref.~[\onlinecite{yakovenko1998influence}] is not gauge-invariant, but when combined with ${\rm Tr}(G_0VG_0V)$ it renders the entire action gauge-invariant. 
Thus, the final result is in agreement with ours, modulo the fact that Ref.~[\onlinecite{yakovenko1998influence}] omits the phonon-only part of the action, the external perturbations $B_x$ and $B_y$, and the anisotropy energy $\gamma$.

The method of Ref.~[\onlinecite{krive1985evidence}] is closer to our work.  Its authors state that ${\rm Tr}(G_0VG_0V)$ does not contribute to the effective action of $\phi$ at leading order and that all the leading contributions come from the chiral anomaly term. This conclusion matches our result. However, the statement that ${\rm Tr}(G_{0}VG_0V)$ is negligible is not substantiated by an explicit calculation in Ref.~[\onlinecite{krive1985evidence}]. As we have seen above, an explicit calculation shows that ${\rm Tr}(G_{0}VG_0V)$ is of the same order as the anomaly term, unless one regularizes it. The role of regularization is not evident in Ref.~[\onlinecite{krive1985evidence}].

Finally, Refs.~\cite{su1988inhomogeneity,ishikawa1988phase} use other regularization methods (inhomogeneity expansion, point splitting) when computing the perturbative terms in the action. As a result, the intermediate steps of the calculation are quite different from ours. 
The end result is nevertheless consistent with ours.

\bibliography{biblio.bib}
\end{document}